\documentclass[conference]{IEEEtran}

\usepackage[T1]{fontenc}
\usepackage{tikz}
\usepackage{graphicx} 
\usepackage{enumitem}
\usepackage{amsmath}
\usepackage{amssymb}
  \usepackage{amsthm}
\usepackage{subfigure}
\usepackage{multirow}
\usepackage{array}
\usepackage{tabularx}
\usepackage{booktabs}
\usepackage{multirow} 
\usepackage{multicol}
\usepackage{fancyhdr}
\usepackage{comment}
\usepackage{makecell}
\usepackage{threeparttable}

\usepackage{algorithm}
\usepackage[noend]{algpseudocode}
\usepackage{listings}
\usepackage{xcolor}
\usepackage{caption}

\lstdefinestyle{customC}{
  language=C,
  basicstyle=\ttfamily\footnotesize,
  frame=lines,
  numbers=left,
  numberstyle=\tiny\color{gray},
  keywordstyle=\color{blue},
  commentstyle=\color{gray},
  stringstyle=\color{orange},
  tabsize=2,
  breaklines=true,
  showstringspaces=false,
  rulecolor=\color{black!30},
}

\DeclareCaptionType{listing}[Listing][List of Listings]

\newcommand*\circled[1]{\tikz[baseline=(char.base)]{
    \node[shape=circle, draw,inner sep=0.5pt] (char) {\footnotesize#1};}}

\renewcommand{\arraystretch}{1.3}

\algnewcommand{\Indent}[1]{\State\hspace{#1}}

\newcommand{\TN}{PathFix} 
\newcommand{\etal}{\textit{et al.}}

\makeatletter
\def\footnoterule{\relax%
  \kern-0pt
  \hbox to \columnwidth{\hfill\vrule width \columnwidth height 0.5pt\hfill}
  \kern3pt}
\makeatother

\begin{document}

\title{\TN{}: Automated Program Repair with Expected Path}

\author{
\IEEEauthorblockN{Xu He}
\IEEEauthorblockA{George Mason University\\
Fairfax, VA, USA\\
xhe6@gmu.edu}
\and
\IEEEauthorblockN{Shu Wang}
\IEEEauthorblockA{Palo Alto Networks Inc\\
Santa Clara, CA, USA\\
shuwang@paloaltonetworks.com}
\and
\IEEEauthorblockN{Kun Sun}
\IEEEauthorblockA{George Mason University\\
Fairfax, VA, USA\\
ksun3@gmu.edu}
}

\maketitle

\begin{abstract}
Automated program repair (APR) techniques are effective in fixing inevitable defects in software, enhancing development efficiency and software robustness. However, due to the difficulty of generating precise specifications, existing APR methods face two main challenges: generating too many plausible patch candidates and overfitting them to partial test cases. To tackle these challenges, we introduce a new APR method named PathFix, which leverages path-sensitive constraints extracted from correct execution paths to generate patches for repairing buggy code. It is based on one observation: if a buggy program is repairable, at least one expected path is supposed to replace the fault path in the patched program. PathFix operates in four main steps. First, it traces fault paths reaching the fault output in the buggy program. Second, it derives expected paths by analyzing the desired correct output on the control flow graph, where an expected path defines how a feasible patch leads to the correct execution. Third, PathFix generates and evaluates patches by solving state constraints along the expected path. Fourth, we validate the correctness of the generated patch.
To further enhance repair performance and mitigate scalability issues introduced by path-sensitive analysis, we integrate a large language model (LLM) into our framework.
Experimental results show that PathFix outperforms existing solutions, particularly in handling complex program structures such as loops and recursion.
\end{abstract}


\section{Introduction}

The software development process often faces inevitable software defects, and a recent study reveals that about 50\% of the software development lifecycle (SDLC) on Linux is consumed to rectify these defects~\cite{Gazzola2019Survey}. As software keeps updating, massive defects continuously emerge. Besides, the manual screening and debugging of defects are excessively time-consuming. Hence, automated program repair (APR) becomes a critical solution to streamline program repair processes, thereby enhancing development efficiency and software robustness~\cite{Alur2018Search,Goues2021APR,Winter2023APRResearch,Hossain2024DeepDive}. 

Automated program repair leverages program test and analysis skills to locate the fault and construct a correctness specification, which subsequently serves as the guidance for generating and validating patches~\cite{Goues2021APR}. Since the specification is either a set of test cases or state constraints, traditional APR methods can be classified into two general categories: test-driven and constraint-driven. Test-driven approaches generate patches by rectifying program behavior to align with the test suite~\cite{Westley2009Genprog,Kim2013PAR}. Constraint-driven approaches extract state constraints from program semantics and derive the patches by solving constraints~\cite{Nguyen2013SemFix,Mechtaev2015DirectFix,Mechtaev2018SemGraft}. With the rise of the large language model (LLM), the APR task has also been packaged as a question-and-answer (QA) task, which has achieved promising performance~\cite{Lutellier2020CoCoNut,Jiang2021CURE,sobania2023analysis}.

However, existing APR methods face two main challenges: imprecise specification and overfitting. First, the imprecise specification leads to a redundant search space in generating patches for faulty programs. Specifications describe correct behaviors. In traditional solutions, the specifications are typically expressed as pairs of input and expected output without a specific path trajectory~\cite{Mechtaev2018SemGraft}. Such a path-insensitive design is a trade-off for scalability. As a result, numerous invalid paths become candidates for rectifying faults. For the LLM, lack of detailed specification in prompts is also the primary reason for the repair failure~\cite{Hossain2024DeepDive,sobania2023analysis}. Second, the overfitting problem in APR happens when the generated patch is only correct according to the specification, but neglects the overall correctness of the entire program. One cause of overfitting is incomplete specification, which arises when the specification fails to cover all failure cases. This issue is particularly common in test-driven approaches that rely on manually constructed test suites~\cite {Goues2021APR}. The LLM-based solution encounters a similar problem due to its lack of self-validation. A study~\cite{sobania2023analysis} evaluates the performance of ChatGPT on APR tasks and reports that it could fix 31 out of 40 buggy programs through four rounds of prompting. Meanwhile, our analysis of the 9 failing patches reveals that 7 out of 9 failures are due to overfitting issues.

In this paper, we propose \TN{}, a path-sensitive APR framework that leverages precise constraints derived from correct execution paths to repair faulty programs. Unlike previous path-insensitive approaches, \TN{} employs path-sensitive specifications to improve repair accuracy. However, implementing \TN{} solely with a static analysis framework presents scalability challenges. To address this, we integrate LLM to enhance repair performance, offering three key advantages. First, deriving and solving path-sensitive specifications can be difficult in complex scenarios, such as loops and function calls, but LLMs can effectively mitigate these limitations. Second, the static-analysis-based framework decomposes the APR task and provides precise path constraints, which serve as more explicit prompts, enabling the LLM to follow a structured, step-by-step chain of thought (CoT). Finally, the validation step in \TN{} compensates for the lack of self-validation in LLMs, helping to prevent overfitting.

From the perspective of the static-analysis-based repair framework, \TN{} consists of four components, i.e., \textit{fault path identification}, \textit{specification inference}, \textit{patch generation}, and \textit{patch validation}. 
First, we conduct the equivalence check between two semantically consistent programs (target and reference). The reference program serves as a correctness judge to determine all faulty paths and provide correct outputs. 
Second, the specification inference module derives the expected paths and extracts the path constraints. 
The expected path refers to the path trajectory that starts from the input conditions, extends through the patch, and arrives at the correct output. 
We argue that after applying an appropriate patch, at least one expected path can be found to replace the fault path to reach the correct output. 
Given the bug position and the correct output, we can explore potential expected paths by tracing the control flow graph (CFG).
%
Third, we solve the path constraints and leverage the component-based synthesis (CBS) approach~\cite{Susmit2010Oracle} to generate the patches.
Fourth, we verify the repaired program to prevent overfitting. If any issues arise, we return to the synthesis step with one additional constraint.

From the perspective of LLM, it can improve \TN{} in the specification inference and patch synthesis modules. For the specification inference module, there are two challenges to implementing \TN{}. First, not all potential expected paths are valid. Therefore, we need to examine the accessibility of expected path candidates, effectively filtering out invalid paths. Also, some expected paths, such as paths within loops or multiple function calls, are too complex to summarize and solve the path constraint. To manage this complexity, we slice expected paths and use LLM to summarize constraints by combining the variable states along paths and implicit context logic. For the patch synthesis module, traditional synthesis tools (e.g., CBS) are also incapable of expressing complicated structures, such as function, struct, and compound operation. LLM can extend the representation capability of traditional synthesis tools and synthesize more complex patch expressions.

We first implement the static-analysis-based framework of \TN{}. Also, we exploit the LLM to enhance the implementation of \TN{} for three modules, i.e., path slicing and path pruning modules in specification inference, as well as the patch generation module.
We compare \TN{} with a static analysis-based method Angelix~\cite{Mechtaev2016Angelix} and SemGraft~\cite{Mechtaev2018SemGraft} and LLM-based work~\cite{sobania2023analysis} on the purpose-designed benchmark QuixBugs \cite{Lin2017QuixBugs} and some actual bugs.
By solving more precise path constraints, \TN{} outperforms SemGraft with a 25/40 repair success rate (SemGraft fixes 18/40) on the QuixBugs dataset. After combining with the LLM, \TN{} can solve 12 more cases. For complex program structures, \TN{} with LLM performs better when the defects are located in the recursion and loop body. Meanwhile, all the defects fixed by SemGraft can be fixed more efficiently by \TN{}. 

In summary, we make the following contributions:
\begin{itemize}
    \item We propose a path-sensitive automated program repair framework (PathFix) to formalize precise expected path constraints, thereby increasing the repair efficiency and addressing the overfitting issue.
    \item We exploit the LLM to enhance the implementation of \TN{} to alleviate the scalability and constraint representation issues coming with precise path analysis.
    \item We implement the prototype of \TN{} and evaluate that it outperforms the existing static-analysis-based and LLM-based solutions.
\end{itemize}

\section{Preliminaries}

\begin{figure}[t]
   \centering
   \includegraphics[width=3.3in]{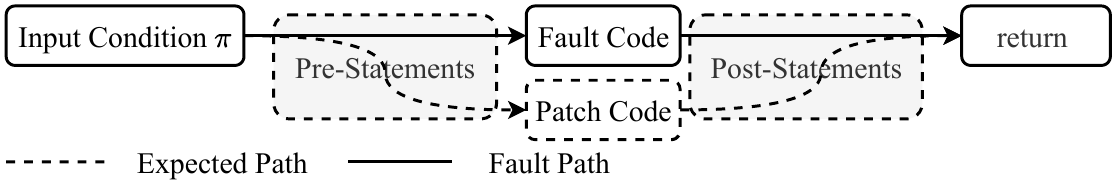}
   \vspace{-0.1in}
   \caption{The expected and fault paths in the target program.}
   \label{fig:ExpectedPath}
   \vspace{-0.1in}
\end{figure}

\subsection{Expected Path}
If fixing a buggy program, the fault paths would be replaced by benign paths, which ensures the patched program behaves correctly. Starting from the input conditions, a benign path goes through the patched code and arrives at the correct output, as illustrated in Figure~\ref{fig:ExpectedPath}. We define these benign paths as expected paths (EP), which is the core idea of this paper. The specifications in previous APR methods only have input and corresponding expected output, while the specific trajectory of the expected path is unknown. This leads to inefficient and inaccurate patch generation.

To generate a correct patch, we search for possible trajectories and express them as path constraints, in which we hypothetically insert a nondeterministic patch predicate. 
By solving the expected path's existence constraint, we can instantiate the predicate and obtain a usable patch. For each fault path, there should exist at least one expected path to replace it. 
Notably, there may exist more than one fault path in a buggy program. Correspondingly, we can generate multiple expected paths to replace them. A valid patch code should be able to satisfy all expected paths' existence constraints simultaneously.
The workflow is illustrated in Sections~\ref{sec:overview} and \ref{sec:design}.

\subsection{LLM enhanced Static Analysis and Program Repair}

Leveraging extensive training data and strong semantic understanding, LLMs have demonstrated promising performance in code summarization and generation tasks. Several studies have explored using LLMs in APR task~\cite{Lutellier2020CoCoNut,Jiang2021CURE,sobania2023analysis,Hossain2024DeepDive}. Most of these approaches treat APR as an end-to-end question-answering (QA) task, where the buggy program is provided as input, and the LLM directly generates a patched version as output.

Despite this progress, APR is inherently a multi-step process. LLMs also exhibit several limitations, including hallucination, randomness, and an inability to self-verify~\cite{Hossain2024DeepDive}, making such a one-hop system unreliable for ensuring patch correctness. In contrast, traditional APR methods typically follow a structured, step-by-step repair framework based on static analysis, which provides a logical reasoning chain to improve repair accuracy. At the same time, LLMs can complement static analysis-based approaches by addressing scalability challenges and assisting in complex code generation tasks~\cite{pei2023variant,Li2024llift}, issues commonly affecting static analysis-based APR solutions.

To leverage the strengths of both paradigms, our work integrates an LLM to enhance multiple components within our static analysis-based APR framework, improving both repair accuracy and efficiency.

\section{A Motivating Example}
\label{sec:overview}

We present a motivating example to illustrate existing methods' limitations and outline the workflow of our proposed approach. Listing~\ref{listing:1} shows a faulty target program, which is a binary search function to search for an element \texttt{x} in a sorted array \texttt{a[]} and return its index if found, or \texttt{-1} if not present. However, line 12 introduces a bug that disrupts the control flow and results in incorrect outputs. 



\begin{lstlisting}[
  style=customC,
  firstline=1,
  lastline=16,
  caption={Motivating Example: binary search program.},
  label={listing:1}
]
    int bin_search(int x, int a[], int len) {
        int L = 0;
        int R = len - 1;
        int m;
        do{ m = (L+R)/2;
            if(x == a[m])
                return m;
            else if(x > a[m]) 
                L = m + 1;
            else 
                R = m - 1;
        } while(L <= m); //(L <= R)
        return -1;
    }   
\end{lstlisting}

\subsection{Limitations of Existing Solutions}
\label{sec:existing}

We apply two existing APR tools, namely, one test-driven method called Angelix~\cite{Mechtaev2016Angelix} and one constraint-driven method called SemGraft~\cite{Mechtaev2018SemGraft}, on this motivating example.
We select these two methods since they are the state-of-the-art solutions that outperform previous well-known methods like Genprog~\cite{Westley2009Genprog}, AE~\cite{Mechtaev2016Angelix}, and SPR~\cite{Long2015SPR}. 


\noindent {\bf Repairs by Angelix}. 
Since Angelix requires test cases to trigger the bug, we first generate two test cases:
(1) $\left \{x = 3, a[] = [1,2,3], length = 3 \right \}$ and (2) $\left \{x = 0, a[] =[1 \right. $  
$\left. ,2,3], length = 3 \right \}$.
We observe that Angelix generates an incorrect patch (i.e., $L<2$) when only case 1 is provided. However, even when both cases are provided, Angelix produces another incorrect patch (i.e., $0 < m$). Notably, Angelix can only generate patches that pass the given test cases. To alleviate this overfitting problem, Angelix depends on the quality of additional test cases to produce a correct patch.

\noindent {\bf Repairs by SemGraft.}
SemGraft first uses the reference program to build the specification and the verification condition (VC).
We select a linear search algorithm (Listing~\ref{listing:2} in Appendix) as the correct reference program to fix this buggy binary search program.
We observe that SemGraft also generates the incorrect patch (i.e., $0 < m$). Note that the VC in SemGraft only focuses on fault paths, and it does not verify the original benign paths after generating the patch. Thus, SemGraft also suffers from the overfitting issue. Besides, since the bug appears in the loop, SemGraft needs to contain the patch's state in each iteration; however, multiple iterations may lead to inefficient repair performance.

\subsection{\TN{} Workflow}
\label{sec:overview1}
We also use Listing~\ref{listing:1} as the target program to illustrate the workflow of \TN{}.
Motivated by SemGraft, we also employ the reference program to ensure we can collect all fault paths for the target program. If the reference program is unavailable, \TN{} can also detect fault paths based on test cases provided by users or other tools, such as Fuzzing or symbolic execution (SE) tool.
We first compare the equivalence between the target and reference programs. The comparison is under the input condition: $\left\{ a\left[0\right]<a\left[1\right]<a\left[2\right], length = 3 \right\}$, which simplifies path exploration in SE and aligns with test cases in Section~\ref{sec:existing}. 
The results are presented in Table~\ref{tab:equiv_check}, where the failure cases are grouped into two fault paths (i.e., P3 and P6). The remaining four paths are benign, exhibiting consistent outputs between the two programs.

\begin{table}[t]
\caption{Equivalence checking with symbolic execution.}
\label{tab:equiv_check}
\vspace{-0.15in}
\begin{center}
\resizebox{0.95\linewidth}{!}{
\begin{tabular}{c|c|c|c}
\toprule
Path ID  & Path Input ($\pi$)           & Output ($\theta_{ref}$, $\theta_{tgt}$)   & Fault           \\ \midrule
P1       & $x=a[1]$                     & (1,1)                                     & $\checkmark$     \\ 
P2       & $x=a[0]$                     & (0,0)                                     & $\checkmark$     \\ 
P3       & $x=a[2]$                     & (2,-1)                                    & $\times$         \\ 
P4       & $x>a[1]\wedge x \ne a[2]$    & (-1,-1)                                   & $\checkmark$     \\ 
P5       & $x<a[1]\wedge x > a[0]$      & (-1,-1)                                   & $\checkmark$     \\ 
P6       & $x<a[0]$                     & (-1,timeout)                              & $\times$         \\ \bottomrule
\end{tabular}
}
\end{center}
\end{table}

\begin{figure*}[t]
    \centering
    \includegraphics[width=\linewidth]{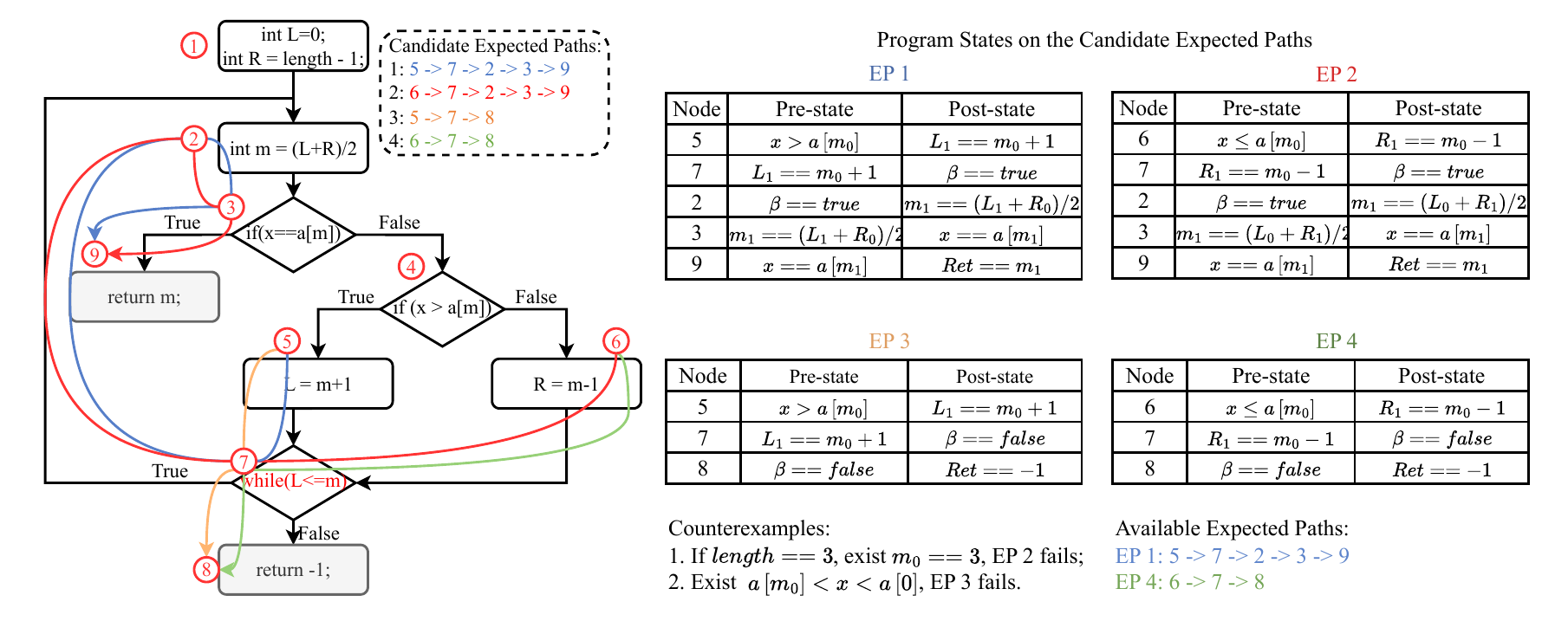}
    \caption{Specification inference (the expected paths and variable states in path constraints).}
    \label{fig:CFG_BinSearch}
   \vspace{-0.2in}
\end{figure*}

After identifying the fault paths, we proceed to derive the expected paths (EPs), which lead to the correct output by deploying an appropriate patch. 
Within the control flow graph (CFG), \TN{} can analyze all possible trajectories between the fault point and all the program exit points, i.e., return statements. As shown in Figure~\ref{fig:CFG_BinSearch}, the CFG indicates four candidate paths, highlighted with different colors.
To simplify the complexity of the loop, we cut the candidate paths to retain only the non-loop fragments between the fault points (node \circled{7}) and exit points (node \circled{8} or \circled{9}) in the last round. 
Notably, the nodes \circled{5} and \circled{6} are included since they represent two distinct inputs for node \circled{7}, providing different pre-execution states.
Among these four EP candidates, EP 1 (blue) and EP 2 (red) are potential alternatives for the fault path P3 in Table~\ref{tab:equiv_check}, as the reference program's output (\texttt{2}) implies a correct program is expected to exit from node \circled{9}. Similarly, EP 3 (yellow) and EP 4 (green) are candidates for fixing the fault path P6.

For each candidate path, we formalize the variable state of each relevant statement and present the states as pre- and post-state pairs, as shown in four tables of Figure~\ref{fig:CFG_BinSearch}. Here, $\beta$ denotes a nondeterministic patch predicate to be solved. The subscript of variables (e.g., $L_1$, $m_1$) indicates the number of times they have been modified along the execution path, e.g., $m_1$ indicates the $m$ value after the first modification.
Hence, a path constraint is formed by the conjunction of states along the expected path.
In essence, generating a patch to fix a fault means finding a solution that satisfies all path constraints.
However, solving the constraints directly is challenging since not all candidate paths are valid. Therefore, we must prune invalid candidate paths.

We prune invalid EPs based on path reachability by examining whether variable states on the candidate path can be satisfied. 
We utilize the expected return value to infer unsatisfiable variable states.
For instance, EP 2 requires $m_0=3$ at node \circled{6} to return value 2, but this violates the array bounds constraint $0 \le m \le length-1$ since $length = 3$.
Thus, accessing $a[m]$ will lead to an out-of-bounds error.
Similarly, if EP 3 is valid, a scenario may arise where $a[m_0] < x < a[0]$, violating the constraint that the array is sorted. Hence, we exclude these two expected paths from candidates since they lead to invalid states.

After filtering invalid paths, we summarize path constraints for valid EP 1 and 4. That is the conjunction of all pre and post-states along the expected path. Then, we transform the conjunction into an implication with the patch predicate $\beta$ to solve. 

\noindent{Given instantiated input constraints:} 
\begin{equation*}
    \left\{length = 3, a[]=\{-1,0,1\} \right\}
\end{equation*}
\noindent{We can infer expected path constraints:}
\vspace{-0.05in}
\begin{equation} \label{equ:syn}
    \begin{array}{ll}
    \exists \beta=f(L, m, R, x, a[], length), \\
    \begin{aligned}
    \text{\bf EP 1:} \ & x=1 \wedge x>a[m_0] \wedge L_1=m_0+1 \wedge \beta= True \\
        & \wedge m_1=(L_1+R_0)/2 \wedge x=a[m_1] \rightarrow m_1=2  
    \end{aligned} \\
    \vspace{0.1in}
    \begin{aligned}
    \text{\bf EP 4:} \   & \ x=-2 \wedge x \leq a[m_0] \wedge R_1=m_0-1 \wedge \beta= false \\ 
        & \rightarrow True
    \end{aligned}
\vspace{-0.1in}
    \end{array}
\end{equation}

In the path constraints, we instantiate $x$, $length$, and $a[]$ to simplify the solving process. To solve the path constraints, we adopt the z3 solver~\cite{Z3-doc} and a component-based synthesis approach~\cite{Susmit2010Oracle}, which selects suitable variables and operators to form a concrete expression $\beta = f(L, m, R, x, a[], length)$ that meets the constraints.

We optimize the component-based synthesizer by prioritizing components (variables and operators).
For example, in the first round, we provide $L$ and $m$, along with logic operators, to the synthesizer, considering their involvement in the original defect. The synthesizer generates an expression: $L > m$; however, the expression fails when verifying along the fault paths.
In the second round, we input $L$, $m$, and $R$ into the synthesizer. The synthesizer generates an expression $L = R$, which successfully passes the verification on the fault paths.

We integrate LLM to address the uncertainties when pruning the expected path, summarizing constraints, and generating patches. When pruning the expected path, we need context information to determine if the path induces potential conflicts. LLM can supplement the program's extra constraint when summarizing the path constraint, such as the loop invariant. When synthesizing the patch, LLM can generate more complicated expressions, such as the function and compound operation. We leverage an LLM to address this issue. For all three tasks, our experiments indicate that LLMs provide insightful intuitive analysis, thereby yielding accurate intermediate and final results.

\section{System Design}
\label{sec:design}

\begin{figure*}[t]
    \centering
    \includegraphics[width=\linewidth]{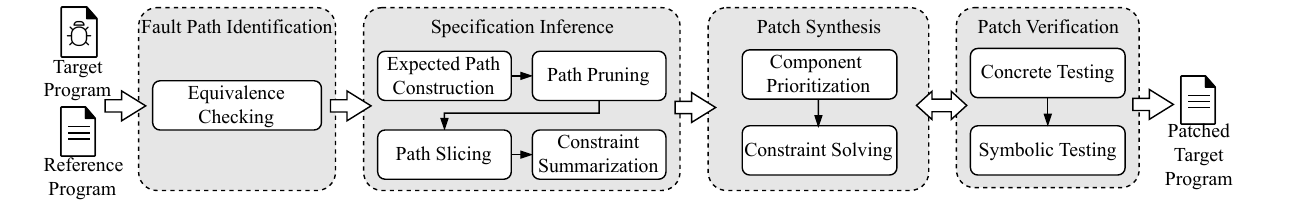}
    \vspace{-0.2in}
    \caption{The static-analysis-based framework of \TN{}. (LLM is plugged into 3 steps: Path Pruning, Constraint Summarization, and Patch Synthesis)} \label{fig:Arch}
    \vspace{-0.15in}
\end{figure*}

In this section, we demonstrate the framework of our \TN{} system. 
As shown in Figure~\ref{fig:Arch}, \TN{} consists of four modules, namely, fault path identification, specification inference, patch synthesis, and patch verification, where the LLM is adopted to enhance the specification inference and patch synthesis modules. 

\subsection{Fault Path Identification}
\label{sec:equivalence}
As demonstrated in Section~\ref{sec:overview1}, we exploit the reference program and symbolic execution (SE) to identify fault paths.
Notably, this step is optional for \TN{}. In practice, if a reference program is unavailable, fault paths can still be identified using test cases provided by users.
%
We design an equivalence-checking method to identify the inconsistent paths between the reference $prog_{ref}$ and target programs $prog_{tar}$. 
We verify the functional equivalence by providing both programs with the same symbolic input and asserting that they produce the same output, i.e., ${assert(prog_{ref}(x)==prog_{tar}(x))}$. For each input condition $\pi$ ($x \in \pi$), the execution follows a path reaching this assertion and checks the output consistency. If a violation of the assertion is detected, the path is labeled with an inconsistency flag $\delta$.
Notably, the equivalence checking method may have limited path exploration under certain scenarios, e.g., the reference target programs might not be strictly semantically equivalent for all input conditions. 
Also, the existing SE tools do not support arrays of unknown length to prevent branch exploration in a loop structure~\cite{Zhou2022Ferry}.
To address this issue, we incorporate a user-defined pre-input condition $\phi_{pre}$ to constrain the exploration space for such scenarios.
Finally, the output of the equivalence checking algorithm is a triplet for all paths, including the input conditions $\pi$, the output $\theta^{out}$, and the inconsistency flag $\delta$. The output $\theta^{out}$ comprises the reference output $\theta^{ref}$ and the target output $\theta^{tar}$. The flag $\delta$ can be used to distinguish the benign paths and fault paths, as shown in the third and forth columns of Table~\ref{tab:equiv_check}.

\subsection{Specification Inference}
Then, we extract expected paths and build path constraints through 4 steps.

\noindent{\bf Step 1: Expected Path Construction.} The expected path starts from the input condition $\pi_i$, extends through the patch predicates $\beta$, and arrives at the correct output $\theta^{ref}$. 
If such an expected path exists, we can instantiate a patch predicate $\beta$ to fix the fault; the expected path will substitute the fault path in the repaired program. 
Thus, the program repair is transformed into solving the expected path's existence constraint.

\begin{figure}[t]
    \centering
    \includegraphics[width=3.3in]{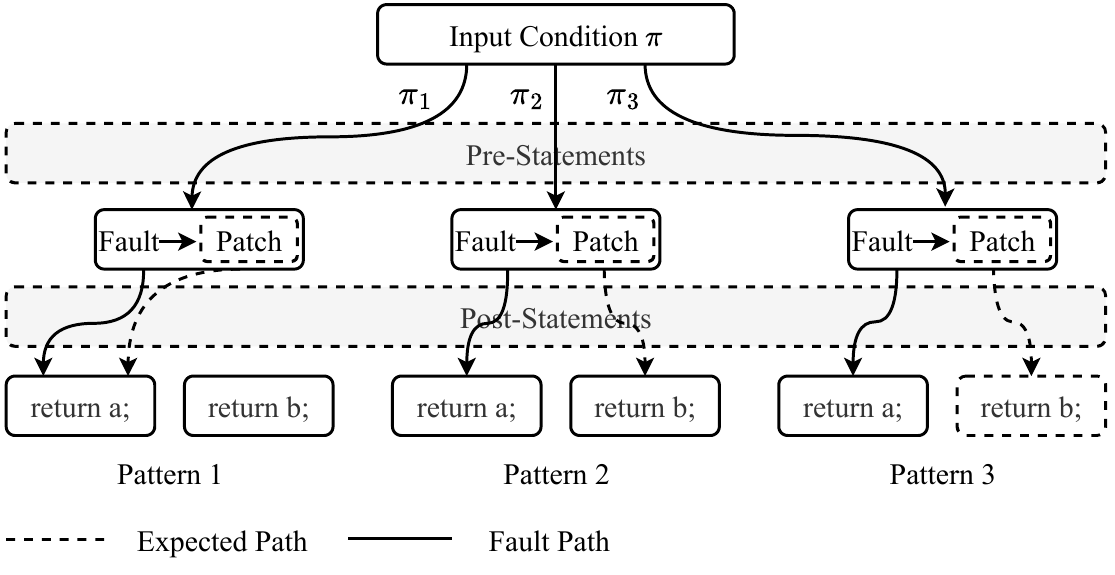}
    \vspace{-0.15in}
    \caption{Three patterns of expected paths in the target program.}
    \label{fig:ExpectedPath3}
    \vspace{-0.15in}
\end{figure}

We first explore all possible expected paths by searching in the control flow graph. 
The expected path can be categorized into three patterns based on the different exit points reached by the expected output and fault output, as illustrated in Figure~\ref{fig:ExpectedPath3}. 
Pattern 1 refers to the expected path reaching the same exit point (but at a different value) as the fault path. In this case, the trajectories of the expected path and the fault path may overlap completely. For example, the bug occurs in sequential statements, such as assignment statements.
Pattern 2 refers to the expected and fault paths reaching different exit points. In this case, the trajectories will only partially overlap. For example, the bug occurs in the if condition or the loop condition, and the program has multiple return statements.
Pattern 3 refers to the expected path reaching an exit point that does not exist before. This indicates that we need to create a new exit point. For example, if the bug causes the program to crash, we can patch it with a return as an error handler.
We determine the expected path's exit point according to the three patterns. 
A particular case arises when the return type of the target program is \texttt{void}. In such cases, we need to identify the implicit termination, typically the leaf node in the control flow graph or the last assignment statement before the leaf node.

We develop a matrix-representation-based path exploration algorithm to derive expected paths automatically.
Specifically, we transform the control flow graph $G_{ctrl}$ into a line graph $L(G_{ctrl})$, in which the nodes correspond to the edges of $G_{ctrl}$, while the edges are converted to nodes.
The line graph is represented as an adjacency matrix, where two nodes are adjacent if their corresponding edges in $G_{ctrl}$ connect to the same node. 
To explore possible expected paths, we integrate the node transition with a depth-first search (DFS) algorithm, starting from the fault location and ending at the expected output.

\noindent{\bf Step 2: Path Slicing.}
The complexity of path constraint based specification can be attributed to the long expected paths, especially multiple iterations of loop procedures, posing challenges for constraint solving. To address this issue, we introduce the concept of \textbf{sliced expected paths} to shorten the path constraint.
A sliced expected path retains a partial path containing the post-statements behind the fault statement (replacing by the patch). If the fault statement occurs in a loop, the sliced expected path only preserves the last round of the loop. Specifically, we slice the expected paths based on two observations.

First, \textit{given a fault path, the fault impacts only the variable states after it is triggered rather than those before it.}
So, we can slice the fault path behind the fault statement, hence narrowing down the scope of states included in the constraint. 
However, if the fault statement is within a loop body, since it is uncertain which iteration triggers the fault, it becomes hard to determine which rounds are impacted by the fault (i.e., bug-related).
Nevertheless, regardless of the iteration in which the fault is triggered, the variable states in the final round must be critical to the fault. Thus, we have the second observation.
\textit{Given an expected path within a loop, if a patch fixes the fault statement, the variable states in the last iteration must be correct.}

The second observation can be inferred using the mathematical induction method.
If the patched target program is proven correct, the intermediate state of the patch is always correct from the $(n-1)^{th}$ iteration to the $n^{th}$ iteration (see detailed proof in Appendix~\ref{app:4}). 
In cases where the fault statement is inside a loop, we only consider the acyclic path in the expected path constraint, i.e., variable states in the last iteration. Also, another reason why we only retain the last iteration is the non-deterministic nature of the expected path. Compared with the fault path, the program may no longer execute along the original path after being fixed. For example, the number of iterations and branches passed may change. 





\noindent{\bf Step 3: Path Constraint Summarization.}
Given an expected path, we model the path constraint as a finite state machine, where each statement's execution corresponds to the state transition between statements. Here, the states refer to the variable states before and after the statement's execution. In this context, we use the Hoare triplet $\{P\}c\{Q\}$~\cite{Hoare1969} to depict the state transition process, 
where $P$ is defined as the pre-execution state, i.e., the variable state before the current statement execution. Similarly, the post-execution state $Q$ represents the variable state after the statement execution. $c$ represents the operation performed by the current statement. Our work redefines operation $c$ as the state transition predicate $C$. For example, supposing operation $c: m=m+1$, we transfer it as a predicate $C: m_1==m_0+1$.
Thus, the modified Hoare triple can be expressed as the following implication.

\begin{equation}
\vspace{-0.05in}
\label{eq:hoare}
    \{P\} c \{Q\} \Leftrightarrow P \wedge C \Rightarrow Q,
\vspace{-0.03in}
\end{equation}
The post-state ($Q$) can be implied by the conjunction of pre-state ($P$) and transfer predicate ($C$). This is formally different from the original inference rule in Hoare logic. 
We depict detailed proof that such a transition is equivalent to the original rule and how to extend such transitions into sequential execution, conditional execution, and cyclic execution in Appendix~\ref{app:3}. 

Taking the sequential execution ($\operatorname{Seq}$) as an example, we can further extend the implication (Formula~\ref{eq:hoare}) to longer paths. For a path consisting of $M$ statements, its path constraint can be expressed as follows.
\begin{equation}
\vspace{-0.05in}
    P \wedge \underset{i \in [1, M]}{\bigwedge} C_i \Rightarrow Q
\vspace{-0.05in}
\end{equation}
Recall that, the fault statement needs to be replaced by a patch to ensure the correctness of the execution path. Hence, we can transform the implication into an existence constraint.
\vspace{-0.05in}
\begin{equation} 
\label{eq:path exist}
    \exists [C_t/\beta], P \wedge \beta \wedge \underset{i \in [1, M], i \neq t}{\bigwedge} C_i \Rightarrow Q
\vspace{-0.05in}
\end{equation}
Here, existence constraints define the objective of solving the patch predicates ($\beta$). 
Note that, apart from the fault statement ($C_t$), all other statements ($C_i$) can be instantiated. Therefore, we can simplify these constraints ($i \in [1, M], i \neq t$) and optimize the complexity of the solution.

To formalize such constraints, we denote variable states as properties of nodes in $L(G_{ctrl})$, and each edge represents the execution of a statement.
When we explore the expected paths, the adjacent node properties could be accumulated.
We can track the state transition via matrix multiplication by extracting the adjacency matrix $M$ from the line graph $L(G_{ctrl})$. 
The one-hop jump of nodes can be expressed as $V_{i+1} = V_i \cdot M$. $V_{i+1}$ represents the reachable nodes of $n_i$ after one hop. 
Thus, we model the line graph $L(G_{ctrl})$ as a finite state machine, where the node transition along the expected path can be viewed as accumulating variable states.

We encountered more uncertainty when accumulating these variable states in real cases. For example, there may be extra functions lying on the expected paths. If we dive into these dependency functions, it will bring redundant variable states into the path constraint, potentially leading to solving failures.
Besides, since the expected path is sliced in a loop structure, the path constraint cannot describe the context information, such as the loop variant logic. 
Therefore, we exploit LLM to enhance the constraint summarization, as shown in Figure \ref{fig:prompt2} (in Appendix). LLM can simplify the possible dependency function logic and context logic, thereby generating more comprehensive path constraints.

\noindent{\bf Step 4: Path Pruning.}
As demonstrated in Section~\ref{sec:overview1}, we observe that not all expected paths are valid in the context of the target program. 
Therefore, we need to verify the expected paths and simplify the path constraints. 
To prune invalid paths, we verify their accessibility. We check if the variable states along the paths satisfy the input constraints and the local variable constraints along the data flow. 
We also apply the graph property analysis method to the data flow graph to obtain the accessible variable list for each statement and the context of variable modification. 
The invalid path will be removed when a candidate path violates the data flow constraint. The detailed derivation process is showcased in Section~\ref{sec:overview1}.

In real-world cases, many invalid path candidates cannot be filtered without the global context information. Therefore, we use LLM to boost the path pruning, and one prompt example is displayed in Figure~\ref{fig:prompt1} (in Appendix). 
In addition, another benefit of using LLM is that it can simplify the path constraints by instantiating partial variables based on the input condition and correct output values.
%
Note that the LLM operates path pruning and constraint summarization in the reverse order. This is because the LLM does not need to explicitly generate constraints to determine a path's validity. Moreover, if constraints are generated for invalid paths first, the LLM may misunderstand the program.

\subsection{Patch Synthesis}
\label{sec:syn}
Our synthesizer is customized based on the component-based synthesis (CBS)~\cite{Sumit2011Brahma} method.
Specifically, for each expected path constraint, we instantiate potential expressions that may appear in the patch predicate ($\beta$) and embed them into the constraint. These instantiated constraints then guide the synthesis of patch expressions. The synthesis process iterates until the generated expression satisfies the expected path constraints, ensuring that the patch passes verification.


The original CBS algorithm randomly selects candidate variables and operators for synthesis. To accelerate patch generation, we prioritize accessible variables and potential operators based on four principles:
\begin{itemize}
\vspace{-0.05in}
    \item[(1)] Variables in the original faulty expression are considered first.
    \item[(2)] Variables modified in expected paths take precedence.
    \item[(3)] Variables modified closer to the patch location are preferred.
    \item[(4)] Other variables involved in modifications are also considered. 
\vspace{-0.05in}
\end{itemize}
Operators are categorized into arithmetic, logical, and bitwise types. Heuristics guide operator selection based on patch context, for example, logical operators are preferred in loops or conditional statements.

The algorithm takes as input the instantiated specification, an initial set of variables N, and possible operator types $T_{ops}$. It initializes variable components $C_{var}$ and iterates through synthesis and verification. First, the synthesis function $CBS()$ generates candidate patches. If no valid patch is found, the algorithm expands the candidate variable and operator sets. If a patch is generated, it is verified against the expected path constraints $\phi_{exp}$. If the constraints are met, the synthesized patch $\beta_{impl}$ is returned; otherwise, a counterexample is generated to refine the instance specification. The component lists $(C_{var}, T_{ops})$ are then reset to optimize subsequent synthesis iterations.

CBS is effective in synthesizing primitive expressions and operator combinations. However, it may fail when handling complex components, such as functions. To address this, we incorporate LLM to enhance patch generation, as illustrated in Figure \ref{fig:prompt3} (in Appendix).

\subsection{Patch Verification}

While the synthesized patch satisfies the constraints of the expected paths, two key issues remain. First, the patch may introduce overfitting, failing to preserve the correctness of the original benign paths. Second, the sliced path specification serves as a necessary but insufficient condition for overall program correctness. Therefore, additional verification is required.

To address these issues, our verification process consists of two interconnected components. The first involves symbolic execution on the faulty paths, which is integrated into the patch synthesis process and iteratively collaborates with the synthesizer to refine the patch. The second focuses on verifying the patched program against its original benign paths to prevent overfitting. To improve efficiency, we leverage test cases generated during the equivalence checking phase to validate benign paths. If a test case fails, the synthesis process is re-entered, using the failing case as an additional constraint to guide subsequent patch generation.


\subsection{LLM Enhancement}

Building on the static-analysis-based framework, we integrate LLM to enhance \TN{}’s performance in three key steps: invalid path pruning, constraint summarization, and patch generation, aligned with the design in Figure~\ref{fig:Arch}. We structure these steps into three progressive tasks, where responses from earlier steps inform subsequent ones, facilitating a logical chain of thought. The prompt templates, with placeholders ({}), are adapted for each case. We list the prompt setting for three steps in Appendix~\ref{app:prompt}.

For instance, in Listing~\ref{listing:1}, we provide the function definition and expected paths to identify invalid paths, along with fault path conditions to aid constraint summarization. Additional information, such as function dependencies or struct definitions, is supplied as needed when relevant to the expected paths or patch expression. Comparing LLM’s responses with static analysis results, we observe that LLM reasoning aligns more closely with human intuition. In patch generation, instead of directly solving constraints in a strict format, LLM analyzes the general logic to ensure $\beta$ satisfies the expected path constraints.

However, two challenges arise. First, a simple prompt template does not guarantee consistent performance across different cases. Second, LLM responses often lack a standardized format, sometimes providing unstructured analytical content or generating patch expressions at incorrect locations.
To address these issues, we apply two strategies. First, we insert examples of expected path constraints into the system prompt, leveraging few-shot in-context learning to help LLM recognize patterns and improve constraint summarization. Second, we adjust the model’s temperature setting, which controls randomness and reasoning ability. Lower temperatures (0–0.5) degrade performance, so we retain the default (1.0 in GPT-4o) to ensure high-quality responses.
Additionally, we enforce structured output formatting. Expected paths are stored in JSON, path constraints in Z3 format, and patch expressions in C syntax. This ensures consistency and facilitates further automated processing.

\section{Evaluation}
In this section, our evaluation mainly investigates 3 research questions:
\begin{description}
    \item[RQ1:] How well does \TN{}~perform, compared with the previous method?
    \item[RQ2:] How much enhancement can LLM bring to \TN{} for each steps?
    \item[RQ3:] What are the reasons for the repair success and failure in \TN{}?
\end{description}

\subsection{Experimental Setup}
\noindent{\bf Dataset.} 
We conduct experiments on QuixBugs~\cite{Lin2017QuixBugs} and 10 real bugs in Busybox and Coreutils. 
First, QuixBugs contains 40 programs of classic algorithms.
We choose programs sharing similar functionality to serve as reference programs, e.g., \texttt{mergesort} and \texttt{bucketsort}.
Second, we also evaluate \TN{} under real-world scenarios by collecting 10 bugs from Busybox and GNU Coreutils, which are also evaluated by SemGraft\footnote{SemGraft presents commit IDs and project versions for 12 real bugs; however, we find 2 commit IDs are invalid.}. 

\noindent{\bf Comparison Setting.} 
We compare \TN{} with SemGraft and Angelix. 
SemGraft relies on reference programs and the CBS synthesizer, while Angelix relies on test cases. 
We also compare the success repair rate with the pure LLM-based solution, in which the user feeds the buggy code and fault cases into the LLM and requires the LLM to generate a patch in a one-step conversation. Notably, we use the pass@1 metric to denote the success repair rate for the LLM-based solution.

\noindent{\bf Environment Setting.}
All the experiments are performed on a Ubuntu 22.04 server with an Intel Xeon 2620 CPU at 2.4 GHz and 16 GB RAM. \TN{} is implemented in Python with dependency on existing tools, including Klee, Joern, Z3, and Brahma. Regarding LLM, we adopt the gpt-4o API.

\begin{table}[t]
\vspace{-0.1in}
\centering
\caption{\label{tab:realbug}Repair performance on real-world bugs.} 
\vspace{-0.1in}
\resizebox{\linewidth}{!}{
\begin{tabular}{c|c|c|c|c|c|c}
\toprule
\multicolumn{2}{c|}{Buggy Prog.}     & {Ref. Prog}          & {\TN{}}& {SemGraft} & {Angelix}& {\TN{} w/o Ref.}\\
\midrule
\multirow{4}{*}{BusyBox}   & sed    & sed of GNU sed     & $\checkmark$ & $\checkmark$        & $\checkmark$ & $\checkmark$        \\ \cline{2-7}
                           & sed    & sed of GNU sed     & $\checkmark$            & $\checkmark$        & $\times$& $\checkmark$        \\ \cline{2-7}
                           & sed    & sed of GNU sed     & $\checkmark$            & $\checkmark$        & $\times$& $\checkmark$        \\ \cline{2-7}
                           & sort   & sort of  Coreutils & $\checkmark$            & $\checkmark$        & $\times$& $\checkmark$        \\  \hline
\multirow{6}{*}{Coreutils} & mkdir  & mkdir of Busybox   & $\checkmark$            & $\checkmark$        & $\times$& $\checkmark$        \\ \cline{2-7}
                           & mkfifo & mkfifo of Busybox  & $\checkmark$            & $\checkmark$        & $\times$& $\checkmark$        \\ \cline{2-7}
                           & mknod  & mknod of Busybox   & $\checkmark$            & $\checkmark$        & $\times$& $\checkmark$        \\ \cline{2-7}
                           & copy   & copy of Busybox    & $\checkmark$            & $\checkmark$        & $\checkmark$            & $\checkmark$        \\ \cline{2-7}
                           & md5sum & md5sum of Busybox  & $\checkmark$            & $\checkmark$        & $\checkmark$            & $\checkmark$        \\ \cline{2-7}
                           & cut    & cut of Busybox     & $\checkmark$            & $\checkmark$       & $\times$& $\checkmark$       \\
\bottomrule
\end{tabular}
}
\vspace{-0.1in}
\end{table}

\subsection{Performance Comparison}

\noindent{\bf Comparison on Real Bug Cases.} 
We make comparisons on 10 real programs. As shown in Table~\ref{tab:realbug}, both \TN{} with LLMs and SemGraft repair all bugs successfully, but Angelix only fixes 3 bugs. Notably, we also collect complete test suites for Angelix and \TN{} w/o Ref. The results show that \TN{} can still generate correct expected paths and patches without a reference program.  
Regarding repair efficiency, \TN{} outperforms SemGraft on repair time and the number of paths included in the constraint. \TN{} required 27 minutes to repair these real bugs, while SemGraft required 45 minutes. 
SemGraft claimed it needs to traverse up to 250 paths during specification inference in a buggy program, while \TN{} with LLMs only needs to consider up to 4 paths after path pruning. 
This is because SemGraft needs to consider more intermediate states in its verification condition, especially when inferring $\beta$ of the predicate in multi-loops, while \TN{} only contains the $\beta$ state in the last round.

\noindent{\bf Comparison on Quixbugs.}
We further evaluate \TN{}'s performance on the Quixbugs dataset.
Table~\ref{tab:result1} demonstrates the repair performance comparison on Quixbugs. 
Among the 40 buggy programs, SemGraft only fixes 18 programs. 
Even without LLM enhancement, \TN{} performs significantly better than SemGraft, fixing 25 programs.
For the programs fixed by SemGraft, \TN{} can also fix all of them.
%
The pure LLM solution can fix 31 cases. In 9 failure cases, there are 7 overfitting errors and 2 synthesis errors.
In this context, overfitting errors mean that LLMs generate an executable program, but it can only fix partial fault paths or introduce new faults. Synthesis errors mean that the patched programs generated by LLMs cannot be executed.
For \TN{} with LLMs, we observed that \TN{} performs best compared to all 3 baselines. There are only 3 failure cases due to constraint errors, as listed in Table~\ref{tab:result2}.

In summary, both \TN{} with LLMs and SemGraft can repair real-world bugs, and \TN{} exhibits higher efficiency. With LLM enhancement, \TN{} further outperforms SemGraft and pure LLM-based solutions. Such observations support the idea that more precise constraints can produce more accurate patches. Besides, task decomposition contributes to the performance of LLMs on APR tasks.

\subsection{Enhancement from LLM}
To investigate how LLMs help \TN{} improve the repair performance, we measure the LLM's contributions in the 3 subtasks: path pruning, constraint summarization, and patch synthesis. 
As illustrated in Table~\ref{tab:ablation}, the findings underscore an overall trend of enhanced performance by integrating the 3 steps.

\begin{table}[t]
\centering
\caption{Repair performance on QuixBugs benchmark.}
\vspace{-0.1in}
\label{tab:result1}
\renewcommand{\arraystretch}{0.9}
\resizebox{0.95\linewidth}{!}{
\begin{tabular}{c|c|c|c|c}
\toprule
\multirow{2}{*}{Method} & \multicolumn{4}{c}{Generated Repairs}    \\ \cmidrule{2-5} 
                            & {Fixed} & {Overfitting} & {\makecell{Synthesis \\ Error}} & \makecell{Constraint \\Error} \\ \midrule
SemGraft                    & 18      & 3           & 6                               & 13 \\  \midrule
LLM                         & 31        & 7           & 2                               & 0  \\  \midrule
\TN{} w/o LLM               & 25      & 0           & 10                              & 5  \\  \midrule
\TN{} w/ LLM                & 37      & 0           & 0                               & 3 
\\ \bottomrule
\end{tabular}
}
\vspace{-0.1in}
\end{table}

First, we observed that LLMs can prune all invalid expected paths of 23 cases. 
As we mentioned in Section \ref{sec:design}, expected path pruning is complicated, especially when the expected path is sliced due to the loop structure. Without LLM, \TN{} can only prune invalid expected paths of 9 cases when the program structure is simple. 
These undetected invalid paths caused multiple iterations or synthesis errors. Even if 100\% coverage cannot be achieved, LLM can also prune partially invalid paths in 15 cases. Compared with \TN{} without LLM, it brings better accuracy and efficiency. 
 
Second, we observed that LLM can successfully summarize precise constraints in 35 cases. We also noticed that even if some invalid expected paths are fed into this step, the LLM may still summarize correct constraints. 
LLM corrects these constraints based on the context information; thereby, generated constraints don't follow the invalid expected path but another valid path, fixing the same fault path. 

Third, we observed that LLM can synthesize correct patches in 37 cases. In contrast, \TN{} can only summarize constraints and synthesize correct patches in 25 cases. The most significant improvement happens in programs involving buggy function calls, as shown in Table \ref{tab:result2}. However, there are still 3 failure cases: 2 failed in path pruning, and 1 failed due to an incorrect path constraint. 

In summary, LLM can enhance all 3 steps in \TN{} and contribute to the overall performance. 
We noticed that LLM is a resilient system. Even if some invalid expected paths bypass the path pruning, the final patches generated by LLM are still correct. This is because LLM generates the patch based on the expected paths and the context information of the entire buggy program, e.g., the loop or recursion invariant logic.

\begin{table}[t]
\centering
\caption{Comparison of repair performance with and without LLM integration.}
\vspace{-0.1in}
\label{tab:ablation}
\resizebox{\linewidth}{!}{
    \begin{tabular}{c|c|c|c}
    \toprule
    Method              & Path Pruning      & Constraint Summarization      & Patch Synthesis \\  \midrule
    \TN{} w/o LLM       & 9                 & 17                            & 25              \\  \midrule
    \TN{} w/ LLM        & 23                & 35                            & 37            
    \\ \bottomrule
    \end{tabular}
}
\vspace{-0.1in}
\end{table}

\subsection{Success and Failure Reason Analysis}
We first investigate three failure reasons, as shown in Table~\ref{tab:result1}. 
For overfitting, \TN{} avoids plausible patches through verification on all paths; however, SemGraft generates 3 plausible patches by only focusing on the fault path. The pure LLM-based solution fails in 7 cases due to hallucination. 
A concrete example is \texttt{detect\_cycle(Node* node)} (Listing~\ref{listing:3} in Appendix), where SemGraft and pure LLMs generate the same patch \texttt{h==NULL}, which fixes the bug but causes new issues when \texttt{h->successor==NULL}. \TN{} generates a similar candidate initially, but re-generates the correct patch after verification failure.
%
%
The second type of failure occurs when the synthesizer cannot generate a patch due to timing out or failing to converge. 
Limited by component-based synthesizer capabilities, the synthesizer may fail to find suitable expressions after traversing all accessible variables and operators.
Imprecise path constraints caused by unpruned invalid expected paths also contribute to synthesis failures.
\TN{} fails to synthesize on 10 programs, and SemGraft fails on 6 programs. 
These programs typically require complex patch expressions with multiple variables, ternary operators, and built-in functions.
Third, no valid constraint can be constructed due to the complex program context. Complex context mainly refers to multi-hop function calls, multi-layer nested loops, and recursive functions. \TN{} fails on 5 programs, and SemGraft fails on 13 programs. 

\begin{table}[t]
\centering
\caption{Repair performance on different bug positions.\label{tab:result2}}
\vspace{-0.1in}
\resizebox{\linewidth}{!}{
\begin{tabular}{l|l|cc|cc|cc}
\toprule
\multicolumn{2}{c|}{\multirow{2}{*}{\bf Defect Position}}   & \multicolumn{2}{c|}{\bf \TN{} w/o LLM}                       & \multicolumn{2}{c|}{\bf SemGraft}  & \multicolumn{2}{c}{\bf \TN{} w/ LLM} \\ \cmidrule{3-8} 
\multicolumn{2}{c|}{}                                                            & \multicolumn{1}{c|}{Fixed} & {Unfixed}                      & \multicolumn{1}{c|}{Fixed}       & Unfixed    & \multicolumn{1}{c|}{Fixed}       & Unfixed  \\  \midrule
\multirow{4}{*}{\bf Assignment}         & {sequence body}                 & {3}        & {0}         & {3}           & 0     & 3  & 0        \\
                                        & {loop body}                     & {6}        & {5}         & {4}           & 7     & 10 & 1       \\
                                        & {recursion body}                & {5}        & {3}         & {2}           & 6     & 7  & 1       \\  
                                        & if branch in loop               & 1          & 0           & 1             & 0     & 1 & 0        \\ \midrule
\multirow{5}{*}{\bf Condition}              & {for condition}             & {3}        & {0}         & {3}           & 0     & 3 & 0        \\
                                        & {while condition}               & {3}        & {0}         & {3}           & 0     & 2 & 1      \\
                                        & {if condition}                  & {0}        & {1}         & {0}           & 1     & 1 & 0      \\
                                        & {if condition in loop}          & {3}        & {1}         & {2}           & 2     & 4 & 0       \\
                                        & {if condition in recursion}     & {1}        & {0}         & {0}           & 1     & 1 & 0       \\  \midrule
\multicolumn{1}{l|}{\bf Function}           & {recursion entry}           & {0}        & {5}         & {0}           & 5     & 5 & 0        \\
\bottomrule
\end{tabular}
}
\vspace{-0.1in}
\end{table}

We further analyze the impact of defect type and location on repair performance. 
As shown in Table~\ref{tab:result2}, most defects in QuixBugs are located in loops or recursion structures.
The main reason for SemGraft's high failure rate in loop bodies is state explosion. 
SemGraft considers patch ($\beta$) states in each loop round, causing high constraint-solving overhead.
For instance, \texttt{shortest\_path\_lengths} (Listing~\ref{listing:5} in Appendix) contains a three-layer loop that may reach 125 rounds with a 5-node input graph, resulting in 125 ($\beta_{1-125}$) states in SemGraft's constraint.
\TN{} alleviates this complexity by constructing sliced expected paths that intercept the last loop round.
For recursion, \TN{} retains the first self-invocation execution flow to construct sliced expected paths.
However, neither \TN{} nor SemGraft can generate correct patches at recursion entry. 
Recursion entry refers to faults in input parameters of recursive functions. 
Such failures indicate that improving path pruning and constraint summarization is urgent for \TN{}.

\TN{} with LLMs significantly improves repair performance for challenging locations like loop and recursion bodies. When expected paths are sliced, path constraints may be under-approximated. LLMs can supplement constraints according to loop logic during path constraint summarization. For recursive entry, LLMs enable function entry representation, which is impossible with component-based synthesis methods.
However, LLMs may make mistakes when handling programs with similar function names and implementations. For example, one failure occurs when processing \texttt{find\_first\_in\_sort()} (Listing~\ref{listing:6}), whose implementation is similar to \texttt{find\_in\_sort()} (Listing~\ref{listing:4}). LLMs construct incorrect constraints similar to those in \texttt{find\_in\_sort()} when they should consider duplicate elements.

\section{Discussion}
\TN{} initially assumes that the reference program is provided by the user, as required in \cite{Mechtaev2018SemGraft}. However, our evaluation demonstrates that a reference program is not strictly required. Instead, test cases can be used to infer fault paths, shifting the responsibility to users to provide comprehensive test coverage. We also explored leveraging LLMs to generate reference programs. For the QuixBugs dataset, the LLM achieved 75\% accuracy (30 out of 40), which is insufficient for reliable use since the reference program must be entirely correct. Additionally, using LLMs to detect fault paths without a reference program failed to guarantee complete path coverage. Future work will focus on designing a LLM agent to generate a reference program.

Another limitation is that the sliced expected path represents an upper approximation of the path constraint. The constructed path constraints form a superset of the expected paths, disregarding the influence of preceding statements. As a result, the synthesis process may fail to converge. To address this, we plan to incorporate loop invariants to extend path constraint coverage. Prior research~\cite{pei2023variant} has shown that LLMs are effective in summarizing invariants, which could further enhance our approach.

\section{Related Work}
\noindent{\bf Test-Driven Approaches.}
Test-driven approaches rely on test suites to form correctness criteria, guiding fault localization, patch generation, and verification, such as GenProg~\cite{Westley2009Genprog}, AE~\cite{Weimer2013AE}, RSRepair~\cite{Yuhua2014RSRepair}, and ACS~\cite{Yingfei2017ACS}.
GenProg~\cite{Westley2009Genprog,Goues2012GenProg(2)} is a landmark work. It uses genetic programming to search for repairs in buggy programs. 
%
The performance bottleneck lies in the efficiency of the search algorithm and precise coverage of the candidate space, inspiring continuous improvement in subsequent work~\cite{Yuhua2014RSRepair,Weimer2013AE}.
RSRepair~\cite{Yuhua2014RSRepair} shows random search outperforms genetic algorithms in patch search. 
AE~\cite{Weimer2013AE} optimizes the search space by merging semantically consistent but syntactically different mutations.
APR tools are also integrated into CI/CD pipelines in industry, e.g., Meta's SapFix tool~\cite{Marginean2019SapFix}, which achieves good performance leveraging high-quality internal codebases.


\noindent{\bf Semantics-Driven Approaches.} 
Semantics-driven methods infer constraint-based specifications that formally cover fault conditions~\cite{Goues2021APR}. 
The key insight lies in constructing more precise constraints. One straightforward way is to summarize constraints from test cases. Semfix~\cite{Nguyen2013SemFix} formulates given test cases into repair constraints using symbolic execution.
Another approach builds constraints based on user experience~\cite{Mechtaev2015DirectFix,Kim2013PAR,Huang2019property}. 
DirectFix~\cite{Mechtaev2015DirectFix} proposes that the best repair involves the least changes and optimizes the patch by minimizing changes. 
Property-based APR~\cite{Huang2019property} summarizes specific safety properties for different bug types.
%
The third way is to learn from correct equivalent code, which provides a functional reference.
Relifix~\cite{Tan2015relifix} utilizes previous code to repair bugs in a new version; however, it fails to cover new features due to its reliance on syntactic similarity.
SemGraft~\cite{Mechtaev2018SemGraft} extends the similarity of reference program from syntactic consistency to semantic consistency.


\noindent{\bf AI-Driven Approaches.}
With the rise of large language models (LLMs), AI-based APR has shown promising results~\cite{Lutellier2020CoCoNut,Jiang2021CURE,sobania2023analysis,Hossain2024DeepDive}.
Coconut~\cite{Lutellier2020CoCoNut} first applies a machine translation architecture to build the mapping from buggy code to benign code.
Cure~\cite{Jiang2021CURE} improves performance by fine-tuning Generative Pre-trained Transformer (GPT)~\cite{radford2018improving}. 
Commercial services like Codex~\cite{OpenaiCodex} and ChatGPT~\cite{OpenaiChatGPT} have also demonstrated outstanding APR performance.
With the improvement of LLM's code generation capability, current LLM-based code repair focuses on prompt engineering to improve repair capabilities.
Misu \etal~\cite{Misu2024Dafny} exploited LLMs to boost formal verification based on Dafny language, which is usually used in APR tasks. 
However, we argue that AI-based APR is not ready to completely replace test-driven and semantics-driven approaches.
First, code generated by LLMs, including patches, still relies on human validation due to the unreliable nature of LLMs. 
Second, AI models are limited in context length, hindering large project repairs.

\section{Conclusion}
In this work, we introduced \TN{}, a novel automated program repair framework designed to address two significant challenges in existing APR methods: imprecise specification and overfitting issues. By leveraging static analysis techniques, including symbolic execution and constraint-solving, \TN{} formalizes expected path constraints, enabling more precise patch generation. We integrate LLMs to improve scalability and constraint handling in \TN{}, facilitating repair performance in complex bug scenarios such as loops and recursions. The evaluation demonstrates \TN{}'s superiority over existing static analysis-based and LLM-based solutions, highlighting its potential to improve automated program repair processes and promote more reliable software development practices. 

\section{Acknowledgment}
During the writing and revision process, we used ChatGPT (model: GPT-4o) to reduce grammatical and syntactical issues and improve the compliance of academic expression standards based on the original content. We thank all reviewers for their valuable suggestions during the submission and review process. This work is partially supported by the US Office of Naval Research grant N00014-23-1-2122.
\clearpage
\bibliographystyle{IEEEtran}
\bibliography{Reference}
\clearpage
\section{Appendix}

\subsection{Code Cases}

\begin{lstlisting}[
  style=customC,
  firstline=1,
  lastline=10,
  caption={Linear search.},
  label={listing:2}
]
    int lin_search(int x, int a[], int len) 
    {
        int m;
        for(m = 0; m < len; m++) {
            if(x == a[m]){
                return m;
            }
        }
        return -1;
    }
\end{lstlisting}

\begin{lstlisting}[
  style=customC,
  firstline=1,
  lastline=15,
  caption={detect cycle.},
  label={listing:3}
]
    bool detect_cycle(Node* node) {
      Node* h = node;
      Node* t = node;
      while (true) {
    // fix: ...||h==NULL
    // Plausible fix: h==NULL
        if (h->successor == NULL)
          return false;
        t = t->successor;
        h = h->successor->successor;
        if (h == t)
          return true;
      }
    }
\end{lstlisting}

\begin{lstlisting}[
  style=customC,
  firstline=1,
  lastline=13,
  caption={find\_in\_sort.},
  label={listing:4}
]
    int find(int arr[],int x,int l,int r)
    {
      if (l == r)
        return -1;
      int m = l + (r - l) / 2;
      if (x < arr[m])
        return find(arr, x, l, m);
      else if (x > arr[m])
        return find(arr, x, m, r);
    // fix: find(arr, x, m+1, r)
      else
        return m;
    }
\end{lstlisting}

\begin{lstlisting}[
  style=customC,
  firstline=4,
  lastline=20,
  caption={find\_first\_in\_sorted.},
  label={listing:6}
]
    int find_first_in_sorted(int arr[], int n, int x) {
        int lo = 0;
        int hi = n;
        while (lo <= hi) { // while (lo < hi)
            int mid = (lo + hi) / 2;
            if (x == arr[mid] && (mid == 0 || x != arr[mid - 1])) {
                return mid;
            }
            else if (x <= arr[mid]) {
                hi = mid;
            }
            else {
                lo = mid + 1;
            }
        }
        return -1;
    }
\end{lstlisting}

\begin{lstlisting}[
  style=customC,
  firstline=9,
  lastline=31,
  caption={shortest\_path\_lengths.},
  label={listing:5}
]
    void shortest_path_lengths(int n, double length_by_edge[MAX_NODES][MAX_NODES]) {
        for (int i = 0; i < n; i++) {
            for (int j = 0; j < n; j++) {
                if (i == j) {
                    length_by_path[i][j] = 0;
                } else if (length_by_edge[i][j] != 0) {
                    length_by_path[i][j] = length_by_edge[i][j];
                } else {
                    length_by_path[i][j] = DBL_MAX;
                }
            }
        }
        double new_dist = 0.0;
        for (int k = 0; k < n; k++) {
            for (int i = 0; i < n; i++) {
                for (int j = 0; j < n; j++) {
                    new_dist = length_by_path[i][k] + length_by_path[j][k]; 
                    // new_dist = length_by_path[i][k] + length_by_path[k][j];
                    length_by_path[i][j] = (new_dist < length_by_path[i][j]) 
                        ? (new_dist) : length_by_path[i][j];
                }
            }
        }
\end{lstlisting}

\subsection{Path Constraint Inference Rules based on Hoare Logic}
\label{app:3}
 
In this paper, we utilize Hoare triple and inference rules to construct the path constraint. The Hoare triple $\{P\} c\{Q\}$ formalizes the execution of a program statement, where $P$ and $Q$ are first-order predicates representing the pre- and post-states, respectively, and $c$ denotes a single statement.
The inference rules extend the execution formalization to the entire program, which consists of various program logic like sequential, conditional, and loop statements. 
The inference rules are based on the substitution rule and backward derivation. Assuming operation $c: m=m+1$, then $\{P\} c\{Q\}$ can be converted into $[m+1/m]Q\{m:m+1\}Q$. This means replacing the variable $m+1$ in state $Q$ with $m$, then the state $P$ is satisfied. On the contrary, if $\{P\} c\{Q\}$ holds true, replacing the variable $m$ in state $P$ with $m+1$ will result in state $Q$.

In our work, we hope to retain all states on the expected path instead of replacing them since any state may be related to patches. So we convert operation $c$ into a state $C$: $m_1 == m_0 + 1$, then we have an implication: ${P} \wedge {C}=>{Q}$. 
This transformation process is equivalent to the inference rules of Hoare logic. Because the substitution rule still holds on the implication.
With the inference rules, we can extend the state transitions to sequential execution, conditional execution, and cyclic execution as implications, as shown in the following.

\noindent{(1) Sequence Structure:}
\begin{equation}
    \frac{\{P\} c_1 \{R\} \, \{R\} c_2 \{Q\}}{\{P\} c_1; c_2 \{Q\}} \Leftrightarrow P \wedge C_1 \wedge C_2 \Rightarrow Q
\end{equation}

\noindent{(2) If-branch Structure:}
\begin{equation}
    \frac{\{P \wedge b\} c_1 \{Q\} \, \{P \wedge \neg b\} c_2 \{Q\}}{\{P\} \operatorname{if}(b)\{c_1\} \operatorname{else}\{ c_2\}\{Q\}} \Leftrightarrow P \wedge (b \wedge C_1 \vee \neg b \wedge C_2) \Rightarrow Q
\end{equation}

\noindent{(3) While-loop Structure:}
\begin{equation}
    \frac{\{P \wedge b\} c\{P\}}{\{P\} \operatorname{while}(b) c \{P \wedge \neg b\}} \Leftrightarrow P \wedge b \wedge C \Rightarrow P \wedge \neg b
\end{equation}

\subsection{Proof of Observation~2}
\label{app:4}
\begin{proof}
Define a proposition $P(n)$: Given an execution path with $n$ iterations, if there is a patch that fixes the fault, all states ($S_n$) in the $n^{th}$ iteration is correct. \\
\textbf{Base Case ($P(1)$)}: Suppose $n = 1$, the current path is equivalent to a loop-free path. Given the patch fixes the bug, $S_1$ should be correct. Hence, $P(1)$ is true.\\
\textbf{Inductive Step ($P(k+1)$)}: Assume $P(k)$ is true for arbitrary positive integer $k$. This means if there is a patch that fixes the fault, then $S_k$ is correct. 
Assuming that the patch has addressed the fault, we need to prove $P(k+1)$. Since $S_k$ is correct, the input for the iteration $(k+1)$ is correct as well, as it is derived from the previous iteration. Therefore, based on the correctness of $S_k$, the state $S_{k+1}$ will also be correct. 
Thus, if $P(k)$ is true, then $P(k + 1)$ is also true. \\
\textbf{Final Step ($P(n)$)}: By the principle of mathematical induction, $p(n)$ is also true, as $k+1$ approaches $n$.
\end{proof}

\subsection{Prompts in LLM-enhanced workflow}
\label{app:prompt}
\begin{figure}[htbp]
    \centering
    \includegraphics[width=0.95\linewidth]{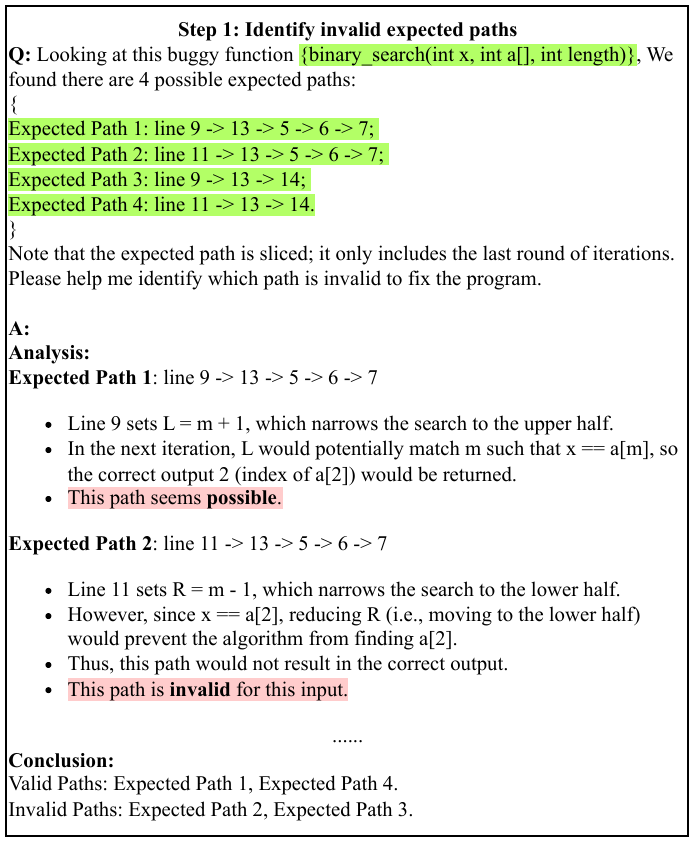}
    \vspace{-0.15in}
    \caption{LLM-enhanced Workflow: Step 1. Identify invalid expected paths.}
    \label{fig:prompt1}
   \vspace{-0.15in}
\end{figure}

\begin{figure}[htbp]
    \centering
    \includegraphics[width=0.95\linewidth]{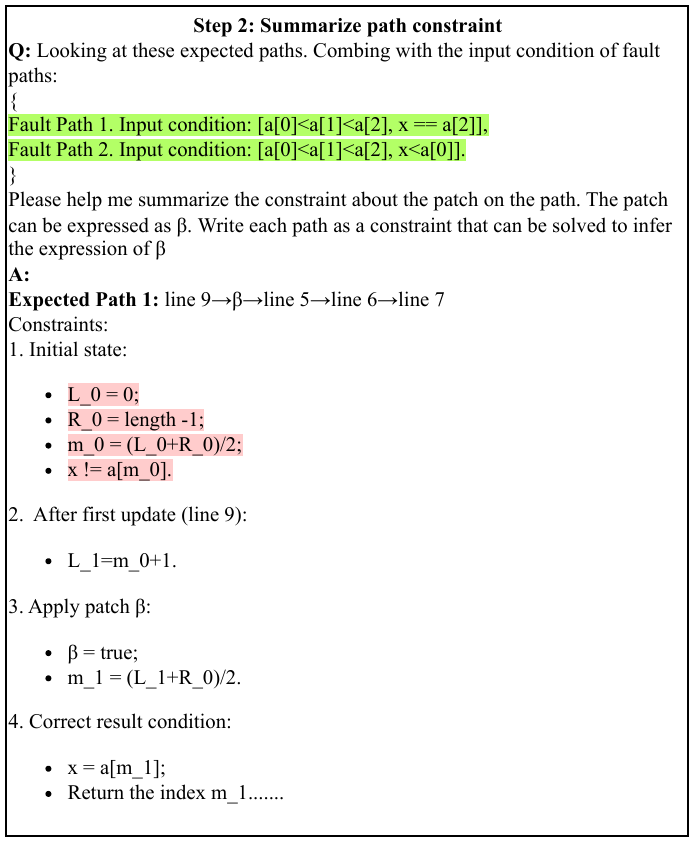}
    \vspace{-0.15in}
    \caption{LLM-enhanced Workflow: Step 1. Identify invalid expected paths.}
    \label{fig:prompt2}
   \vspace{-0.15in}
\end{figure}

\begin{figure}[ht]
    \centering
    \includegraphics[width=0.95\linewidth]{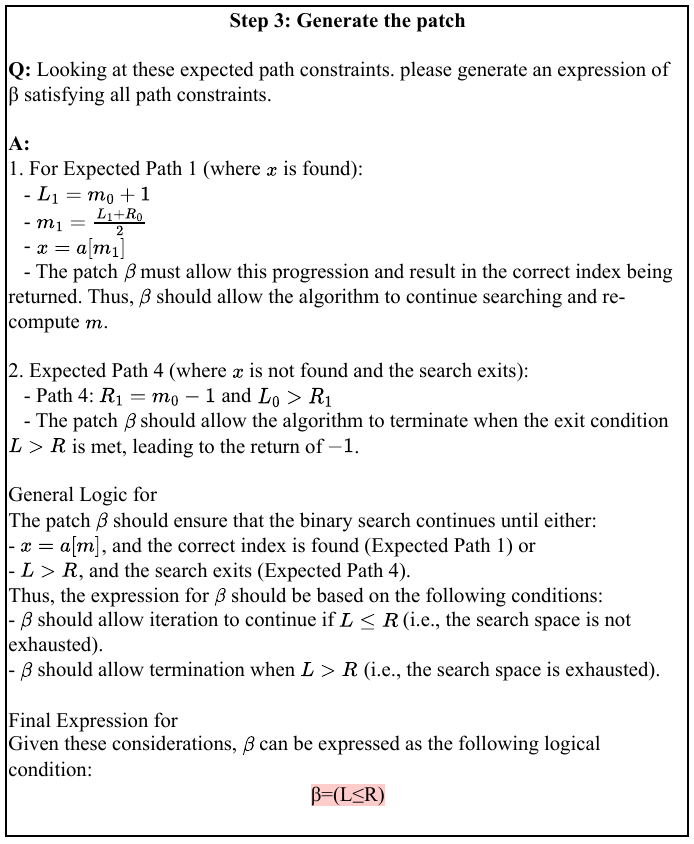}
    \vspace{-0.15in}
    \caption{LLM-enhanced Workflow: Step 1. Identify invalid expected paths.}
    \label{fig:prompt3}
   \vspace{-0.15in}
\end{figure}

\end{document}